\documentclass[a4paper]{jpconf}
\usepackage{graphicx}
\usepackage{amsmath}
\usepackage{amssymb}
\usepackage{bm}
\usepackage{cases}

\usepackage{comment}

\begin{document}
\title{Complex-angle analysis of electromagnetic waves on interfaces}

\author{Daigo Oue$^1$}

\address{$^1$Division of Frontier Materials Science, Osaka University, 1-3 Machikaneyama, Toyonaka, Osaka, Japan 560-0871}

\ead{daigo.oue@gmail.com}

\begin{abstract}
  Electromagnetic wave is reflected and refracted at interfaces, satisfying Fresnel-Snell law which is required by conservations of energy and momentum. If the incident angle is lower than the critical angle, we can use this Fresnel-Snell law, but the Fresnel-Snell law is modified in the case of existence of dissipation ($\tilde{n} = n + i\kappa, \kappa > 0$) or in the condition of total internal reflection. In the cases, we have to extend the angle of refraction from real number to complex number ($\theta \rightarrow \tilde{\theta} = \theta + i\psi, \psi \neq 0$). In this paper, by using complex-angle approach, we analyse the behaviour of electromagnetic waves in various kind of interfaces: dielectric - dielectric system, dissipative dielectric - dielectric system, and metal - dielectric system. We show that iso-frequency curves in wavenumber space is opened in the case where $n, \kappa > 0$, and closed either when $n \rightarrow 0$ or when $\kappa \rightarrow 0$ ('Lifshitz transition' of electromagnetic waves). Excess momentum (wavenumber) and anomalous spin (circular polarization) emerging with the transition are also discussed.
\end{abstract}

\section{Introduction}

\subsection{Fresnel-Snell law}
Snell law (\ref{eq:Snell}) is one of the fundermental laws which we use to analyse the behaviours of electromagnetic waves on interfaces, and which is required by momentum conservation on interfaces \cite{novotny2012principles}.
\begin{align}
  n_i \sin \theta_i = n \sin \theta \label{eq:Snell}
\end{align}
One of the most important thing about Snell law is that once we obtain angle of refraction from Snell law, which gives wavevector as will be explained in the subsection \ref{sbsec:rotation}, we can obtain the amplitude of reflected and refracted waves by using Fresnel coefficients \cite{novotny2012principles}.

\subsection{Rotation for calculation of reflected and refracted waves} \label{sbsec:rotation}
To give explicit forms of reflected and refracted waves, we need to take the propagation directions into consideration.
One method to do so is to rotate the wave vector and the field vector:
\begin{align}
  \bm{k} = \begin{bmatrix} 0 \\ 0 \\ k_0 \end{bmatrix} \rightarrow \bm{k} = R(\Theta) \begin{bmatrix} 0 \\ 0 \\ nk_0 \end{bmatrix}, \bm{E} = \begin{bmatrix} E_p \\ E_s \\ 0 \end{bmatrix} \rightarrow \bm{E} = R(\Theta) \begin{bmatrix} t_p E_p \\ t_s E_s \\ 0 \end{bmatrix}
\end{align}
by using a rotation matrix:
\begin{align}
  R(\Theta) = \begin{bmatrix}
    \cos \Theta & 0 & \sin \Theta \\
    0 & 1 & 0 \\
    -\sin \Theta & 0 & \cos \Theta 
  \end{bmatrix}.
\end{align}
Here, for reflected waves $\Theta = \pi/2 - \theta_i$, whereas for refracted waves $\Theta = \theta$.
Then, we can get the explicit forms by substituting them into $\bm{E}\exp (i\bm{k}\cdot\bm{r})$.

\section{Complex angle}

\subsection{Requirement of complex number for angle of refraction}
We need to introduce imaginary part in the angle of refraction ($\theta \rightarrow \tilde{\theta} = \theta + i\psi$) due to satisfying momentum conservation on interfaces when the refracted waves have decay structure such as the presence of dissipation (or total internal reflection which is discussed in \cite{bekshaev2013mie, bliokh2014extraordinary}).

\subsection{Complex-angle Snell law and complex-angle rotation}
With the complex-angle ($\tilde{\theta} = \theta + i \psi$) and complex-index of refraction ($\tilde{n} = n + i \kappa$), Snell law are modified as
\begin{align}
  n_i \sin \theta_i &= \tilde{n} \sin \tilde{\theta}. \label{eq:complex_Snell}
\end{align}
Note that from the imarginary part of (\ref{eq:complex_Snell}) the imaginary part of the complex-angle is a function of the real part ($\psi = \psi(\theta)$).

\par Complex-angle rotation matrix is given by
\begin{align}
  R\left( \tilde{\theta} = \theta + i\psi \right) = 
  \cosh \psi 
  \begin{bmatrix}
    \cos \theta & 0 & \sin \theta \\
    0 & 1/\cosh \psi & 0 \\
    -\sin \theta & 0 & \cos \theta
  \end{bmatrix} 
  - i \sinh \psi 
  \begin{bmatrix}
    \sin \theta & 0 & -\cos \theta \\
    0 & 0 & 0 \\
    \cos \theta & 0 & \sin \theta
  \end{bmatrix}.
\end{align}
Complex-angle rotation yields wavevector of refracted wave:
\begin{align}
  \tilde{\bm{k}} &= R\left( \tilde{\theta} \right)\begin{bmatrix} 0 \\ 0 \\ \tilde{n}k_0\end{bmatrix} = \bm{k} + i\bm{\eta} \label{eq:wavevector}
\end{align}
where
\begin{align}
  \bm{k} = k_0 \bar{\bm{k}} \equiv k_0 \begin{bmatrix} n \sin \theta \cosh \psi - \kappa \cos \theta \sinh \psi \\ 0 \\ n \cos \theta \cosh \psi + \kappa \sin \theta \sinh \psi \end{bmatrix} ,
  \bm{\eta} = k_0 \bar{\bm{\eta}} \equiv k_0 \begin{bmatrix} n \cos \theta \sinh \psi + \kappa \sin \theta \cosh \psi \\ 0 \\ -n \sin \theta \sinh \psi + \kappa \cos \theta \cosh \psi \end{bmatrix} \label{eq:k_eta}.
\end{align}
Likewise, we obtain the field vector of refracted wave:
\begin{align}
  \bm{E}_0 = R\left(\tilde{\theta}\right)\begin{bmatrix}t_p E_p \\ t_s E_s \\0 \end{bmatrix} = \begin{bmatrix}t_p E_p \left(\cos \theta \cosh \psi - i\sin \theta \sinh \psi \right) \\ t_s E_s \\ -t_p E_p \left(\sin \theta \cosh \psi + i\cos \theta \sinh \psi\right)\end{bmatrix}
  \label{eq:fieldvector}
\end{align}

\section{'Lifshitz transition' of electromagnetic waves}

\subsection{Discontinuous transition of iso-frequency curve}
\begin{figure}[tbp]
  \centering
  \includegraphics[width=6in]{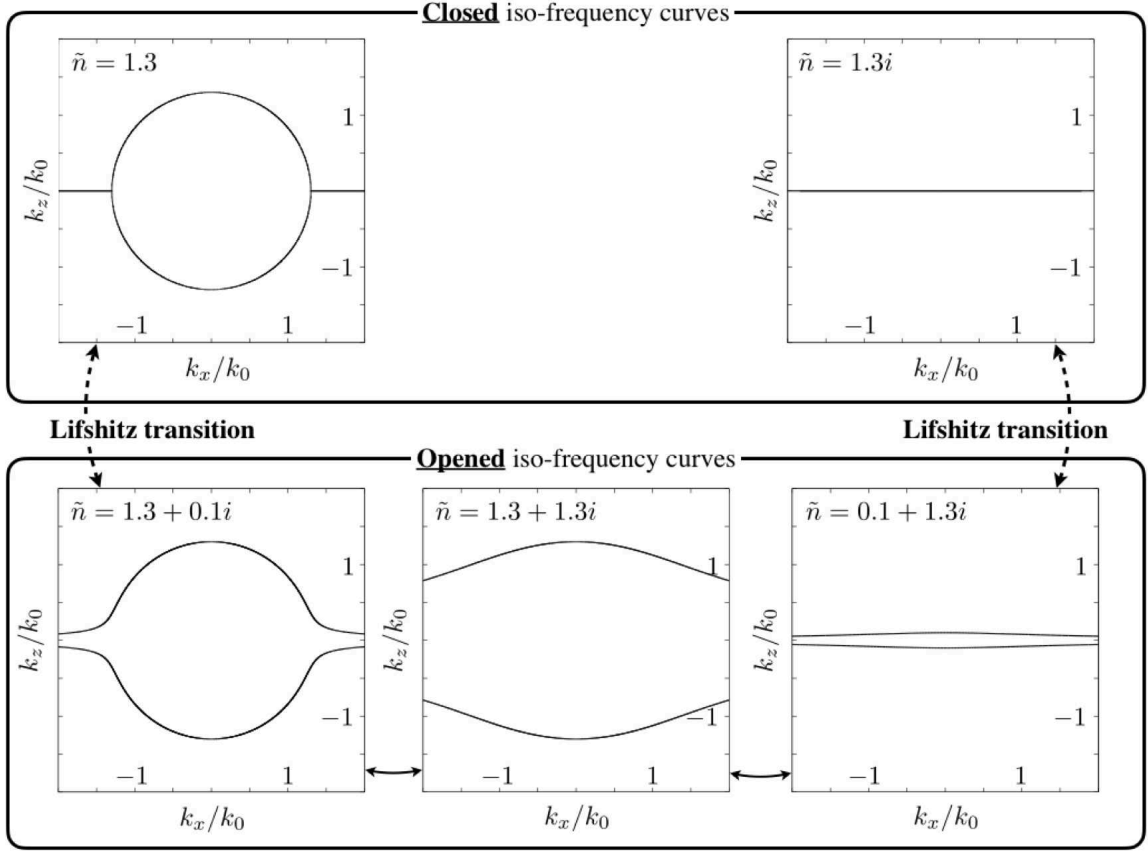}
  \caption{Iso-frequency curves drawn by wavevector for interfaces between various complex refractive indeces $\tilde{n}$ and  vacuum. The iso-frequency curve is closed either when real part vanishes ($\textrm{Re}(\tilde{n}) = n \rightarrow 0$) or when imaginary part ($\textrm{Im}(\tilde{n}) = \kappa \rightarrow 0$) vanishes, while it is opened when both real and imaginary part are non-zero ($\textrm{Re}(\tilde{n}) \times \textrm{Im}(\tilde{n}) = n\kappa \neq 0$).}
  \label{fig:Lifshitz}
\end{figure}

The wavevector (\ref{eq:wavevector}) which we obtain by complex-angle rotation satisfies the complex dispersion relation (\ref{eq:dispersion}), which holds in the dissipative media.
\begin{align}
  \tilde{\bm{k}} \cdot \tilde{\bm{k}} = \frac{\omega}{c}\tilde{n} \label{eq:dispersion}
\end{align}
Figure \ref{fig:Lifshitz} shows iso-frequency curve drawn by wavevector discontinuously changes its shape (open$\leftrightarrow$close transition). We call this 'Lifshitz transition' of electromagnetic waves.

\subsection{Surface mode does not exist}
Once iso-frequency curves are open, surface mode cannot exist on the interfaces (observe that there is no condition of $k_z = 0$ in the bottom graphs in Figure \ref{fig:Lifshitz}). 
Surface modes have pure imarginary wavenumber perpendicular to the interfaces, but by dissipation this pure imaginary component couples out to real (propagating) component, and the modes become reaky (For the rigorous proof, see \cite{supplementary}).

\subsection{Excess momentum and anomalous spin}
\begin{figure}[tbp]
\includegraphics[width=2.5in]{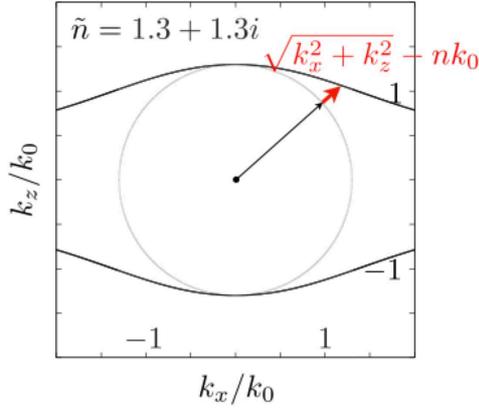}
\begin{minipage}[b]{3.5in}
  \caption{Excess momentum production by dissipation. The solid curve is the iso-frequency curve with the presence of dissipation ($\tilde{n} = 1.3 + 1.3i$), while the grey circle is the iso-frequency curve without the presence of dissipation ($\tilde{n} = 1.3 + 0i$).}
  \label{fig:excess}
\end{minipage}
\end{figure}

In Figure \ref{fig:excess}, The black curve is the iso-frequency curve with the presence of dissipation ($\tilde{n} = 1.3 + 1.3i$), while the grey circle is the iso-frequency curve without dissipation ($\tilde{n} = 1.3 + 0i$). The black curve covers the grey circle. This means that the electromagnetic waves near interfaces with dissipation have larger wavenumber (momentum) than the waves without dissipation:  
\begin{align}
  \left|\textrm{Re} \left( \tilde{\bm{k}} \right) \right| = |\bm{k}| = k_0\sqrt{\left( \bar{k}_x \right)^2 + \left( \bar{k}_z \right)^2} > nk_0,
\end{align}
although dissipation generally changes not the wavenumber but only the intensity profile of waves.
This curious phenomena is caused by the presence both of dissipation and interface (cannot happen in bulk).

\begin{figure}[tbp]
  \begin{minipage}{3.1in}
    \centering
    \includegraphics[width=3.1in]{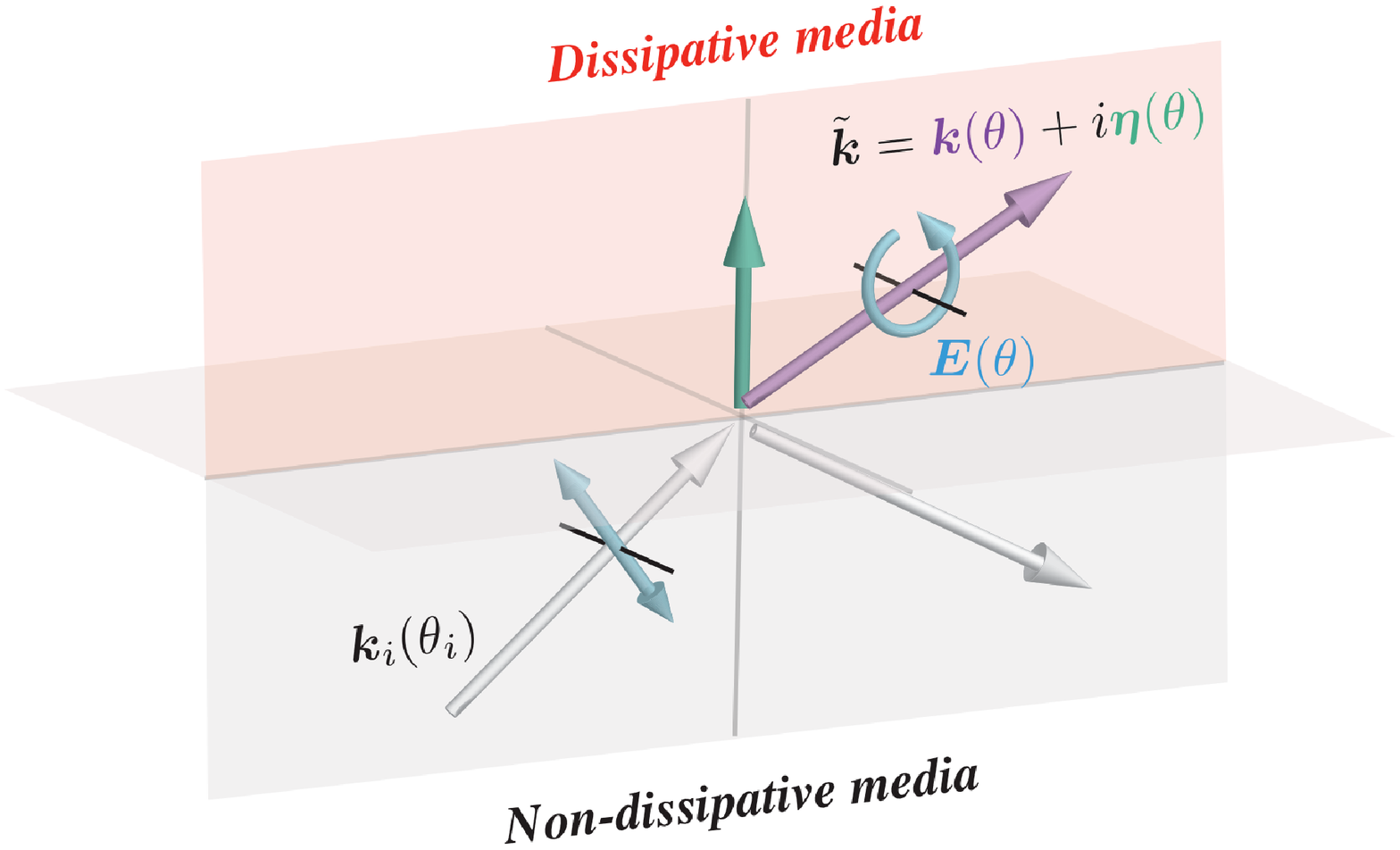}
    \caption{Anomalous spin (p-polarization).}
    \label{fig:p_pol}
  \end{minipage}
  \begin{minipage}{3.1in}
    \centering
    \includegraphics[width=3.1in]{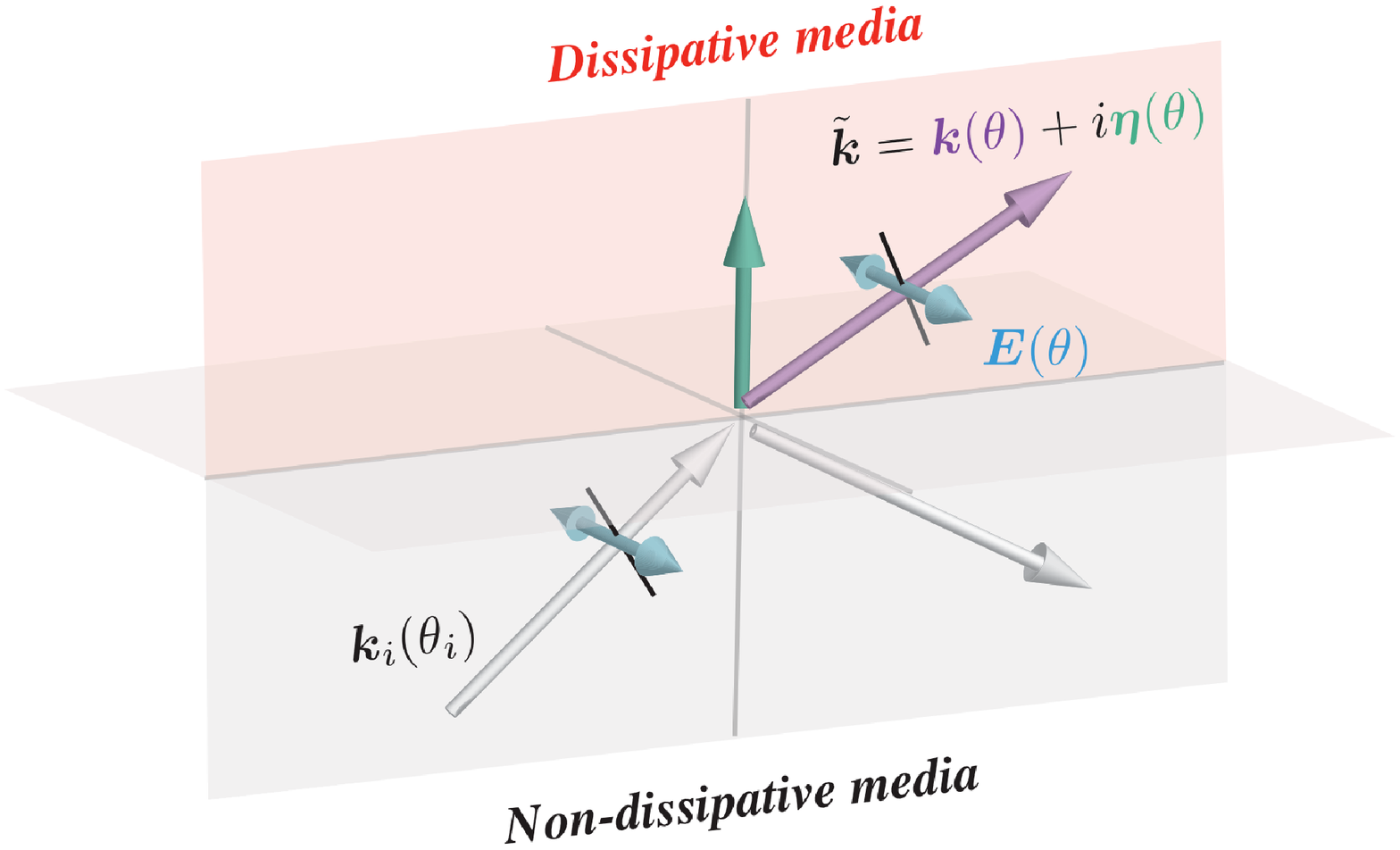}
    \caption{No spin (s-polarization).}
    \label{fig:s_pol}
  \end{minipage}
\end{figure}

Define spin of electromagnetic waves as
\begin{align}
  \bm{s} \equiv \frac{g}{4\omega}\textrm{Im} \left( \bm{E}^\star \times \bm{E} \right).
\end{align}
This spin vector is generally parallel to the wavevector because of Maxwell transversality condition.
Near the dissipative interfaces, however, electromagnetic wave has the longitudinal component, and the spin vector can be non-parallel to the wavevector.
With p-polarization incidence, anomalous spin ($s_y \neq 0$) is generated, while no spin with s-polarization(compare Figure \ref{fig:p_pol} and \ref{fig:s_pol}).

\section{Conclusion}
By utilizing complex-angle approach, we have shown that discontinuous transition of iso-frequency curve ('Lifshitz transition' of electromagnetic waves) occurs when either real part or imaginary part of the complex-index of refraction vanish ($\textrm{Re}(\tilde{n}) \times \textrm{Im}(\tilde{n}) = n\kappa = 0$).
Once the transition occurs, surface mode does not exist on the interfaces.
Also, extraordinary behaviours of electromagnetic waves: excess momentum and anomalous spin emerge.

\section*{References}
\providecommand{\newblock}{}

\end{document}

\title{Complex-angle analysis of electromagnetic waves on interfaces: supplementary material}

\author{Daigo Oue$^1$}

\address{$^1$Division of Frontier Materials Science, Osaka University, 1-3 Machikaneyama, Toyonaka, Osaka, Japan 560-0871}

\ead{daigo.oue@gmail.com}

\begin{comment}
\begin{abstract}
\end{abstract}
\end{comment}

\section{Fresnel coefficients}
We can derive the following four Fresnel coefficients from the continuity condition on the interfaces:
\begin{align}
  r_p &= \frac{-n_i \cos \theta + n \cos \theta_i}{n_i \cos \theta + n \cos \theta_i}, \label{eq:Fresnel_rp} \\
  r_s &= \frac{n_i \cos \theta_i - n \cos \theta}{n_i \cos \theta_i + n \cos \theta}, \label{eq:Fresnel_rs} \\
  t_p &= \frac{2 n_i \cos \theta_i}{n_i \cos \theta_i + n \cos \theta}, \label{eq:Fresnel_tp} \\
  t_s &= \frac{2 n_i \cos \theta_i}{n_i \cos \theta + n \cos \theta_i}. \label{eq:Fresnel_ts}
\end{align}

\section{Complex Snell law}
With complex-index of refraction and the complex-angle, Snell law is modified as
\begin{align}
  n_i \sin \theta_i = \tilde{n} \sin \tilde{\theta},
\end{align}
which is explicitly split into to simultaneous equations:
\begin{subnumcases}
  {}n \sin \theta \cosh \psi - \kappa \cos \theta \sinh \psi = n_i \sin \theta_i, \label{eq:complex_Snell_real}\\
  n \cos \theta \sinh \psi + \kappa \sin \theta \cosh \psi = 0. \label{eq:complex_Snell_imaginary}
\end{subnumcases}

\section{Proof of prohibition of the existence of the surface mode}
In the case where both refractive index and dissipation coefficients are non-zero, surface mode cannot exist but couple out.
For existence of surface mode, it is necessary to satisfy $\bar{k}_z = 0$.
\begin{align}
 \bar{k}_z = n \cos \theta \cosh \psi + \kappa \sin \theta \sinh \psi = 0.\label{eq:condition_surface_mode}
\end{align}
In the condition that $n \neq 0$ and $\theta \neq \pi/2$, (\ref{eq:condition_surface_mode}) is transformed as:
\begin{equation}
  1+\frac{\kappa}{n}\tan \theta \tanh \psi = 0.\label{eq:condition_surface_mode01}
\end{equation}
Substitute Imaginary part of Snell law (\ref{eq:complex_Snell_imaginary}) into this (\ref{eq:condition_surface_mode01}), we get
\begin{align}
  1-\left( \frac{\kappa}{n} \right)^2 \tan^2 \theta &= 0, \notag \\
  \tan \theta &= \pm \frac{n}{\kappa}. \label{eq:condition_surface_mode01_1}
\end{align}
According to (\ref{eq:complex_Snell_imaginary}), when $\theta \rightarrow \tan^{-1} (\pm n/\kappa)$, $\psi \rightarrow \mp \infty$.
On the other hand, (\ref{eq:complex_Snell_real}) can be calculated as below.
\begin{align}
  n_i \sin \theta_i &= n \sin \theta \cosh \psi - \kappa \cos \theta \sinh \psi \notag \\
  \cfrac{n_i \sin \theta_i}{n \cosh \psi} &= \sin \theta - \cfrac{\kappa}{n}\cos \theta \tanh \psi \notag \\
  \intertext{when $\theta \rightarrow \tan^{-1} (\pm n/\kappa) \equiv \pm \theta_{max}$ and $\psi \rightarrow \mp \infty$,}
  0 &= \sin (\pm \theta_{max}) \pm \cfrac{\kappa}{n}\cos \theta_{max} \notag \\
  0 &= \tan (\pm \theta_{max}) \pm \cfrac{\kappa}{n} \notag \\
  0 &= \pm \cfrac{n}{\kappa} \pm \cfrac{\kappa}{n} \notag \\
    &\left(\cfrac{\kappa}{n}\right)^2 + 1 = 0 \label{eq:real_snell_violated}  
\end{align}
However, since $n,\kappa \in \mathbb{R}$ and $n,\kappa \neq 0$, this equation (\ref{eq:real_snell_violated}) is violated. Thus, it can be said that $\theta \neq \tan^{-1} (n/\kappa)$, $\psi \neq \mp \infty$,  and $\textrm{Re} \left(\bar{k}_z\right) \neq 0$. (Equation (\ref{eq:condition_surface_mode}) does not hold, and surface mode does not exist.)